\documentclass{PoS}
\usepackage{dsfont}

\title{Non-perturbative renormalization of flavor singlet quark bilinear operators in lattice QCD}

\ShortTitle{Non-perturbative renormalization of flavor singlet quark bilinear operators in lattice QCD}

\author{Gunnar S. Bali, Sara Collins, Meinulf G\"ockeler, \speaker{Stefano Piemonte}%
         \\
        Universit\"at Regensburg, Institute for Theoretical Physics, D-93040 Regensburg, Germany\\
        E-mail: \email{gunnar.bali@ur.de}, \email{sara.collins@ur.de}, \email{meinulf.goeckeler@ur.de}, \email{stefano.piemonte@ur.de}}

\author{Andr\'{e} Sternbeck\\
        Theoretisch-Physikalisches Institut, Friedrich-Schiller-Universit\"at Jena, 07743 Jena, Germany\\
        E-mail: \email{andre.sternbeck@uni-jena.de}}

\abstract{We report on our studies of the renormalization of flavor singlet quark bilinear operators in lattice QCD. The renormalization constants are determined non-perturbatively using gauge field ensembles with $N_f=2$  dynamical clover improved fermions. The renormalization is performed in the RI'-MOM scheme. The difference between flavor singlet and non-singlet quark bilinear operators is a disconnected contribution, which has to be evaluated by stochastic estimators. We compare our results for the running with perturbation theory.}

\FullConference{34th annual International Symposium on Lattice Field Theory\\
                 24-30 July 2016\\
                 University of Southampton, UK}

\usepackage{subfigure}
\newcommand{\arxiv}[1]{arXiv:\,\href{http://arxiv.org/abs/#1}{{\tt #1}}}
\usepackage{slashed}
\usepackage{amsmath}

\begin{document}

\section{Introduction}

Bare composite operators of quark and gluon fields have to be renormalized in general to provide a well defined continuum limit for hadron matrix elements computed on the lattice. The RI'-MOM scheme is a regularization-independent scheme suitable for computations both in continuum perturbation theory and on the lattice by means of Monte Carlo simulations \cite{MA1}. Flavor non-singlet quark bilinear operators have been successfully renormalized in the RI'-MOM scheme in many previous numerical studies with Wilson and staggered fermions \cite{MG1,MG2,CO1,LY1}.

In this contribution we present the determination of the renormalization constant of flavor singlet quark bilinear operators $Z_{s.}$. The renormalization of matrix elements of phenomenological interest is the main motivation for the non-perturbative study of $Z_{s.}$. The strange quark contribution to the spin of the nucleon is an example of a computation that requires the knowledge of singlet renormalization constants. The spin of the nucleon is the sum of the contribution of the quark spins $\Delta \Sigma$, of the angular momentum of the quarks, $L_q$, and of the spin and total angular momentum of gluons, $\Delta G$,
\begin{equation*}
 \frac{1}{2} = \frac{1}{2} \Delta \Sigma + L_q + \Delta G ~~~~~~~\Delta \Sigma = \Delta u + \Delta d + \Delta s\,,
\end{equation*}
where the contribution of heavy quarks is neglected. The axial singlet renormalization constant is required to compute the contribution $\Delta u + \Delta d$ coming from the spin of the light quarks \cite{SB3}. 

We focus on the two flavor ensembles generated by the QCDSF and RQCD collaborations. A similar calculation of the singlet renormalization constants has been already performed for the $N_f = 3$ flavor case employing the Feynman-Hellmann approach in Ref.~\cite{CH1}. Our preliminary numerical results show that the disconnected contribution to the singlet renormalization constants can reliably be estimated directly, using stochastic noise methods, therefore avoiding the limitations of the Feynman-Hellmann method.

\section{The RI'-MOM scheme for quark bilinears}

The renormalization constants in the various RI'-MOM schemes $Z^\Gamma(\mu)$ are defined by the renormalization condition in the chiral limit
\begin{equation}\label{renocond}
 \frac{1}{12} Z_q^{-1}(\mu) Z^\Gamma(\mu) \textrm{Tr}(\tilde{V}^\Gamma(p) \tilde{V}^\Gamma_{\textrm{Born}}(p)^{-1}) = 1\,,
\end{equation}
where $\tilde{V}^\Gamma(p)$ is the amputated vertex function at a momentum $p$ for the operator 
\begin{equation}
 \Gamma=\{\mathds{1}, \gamma_5,\gamma_\mu,\gamma_\mu\gamma_5,\sigma_{\mu\nu}\}\,.
\end{equation}
The above renormalization condition requires essentially the renormalized vertex function to be equal to its tree level counterpart at the momentum corresponding to the renormalization scale. The renormalization scale is set to be equal to the momentum, i.e. $\mu^2 = p^2$.

The current $\bar{\psi} \Gamma \psi$ is inserted at zero momentum in the vertex function in the RI'-MOM scheme
\begin{equation}\label{vertex}
 V^\Gamma (p) =   \sum_{x,y,z} \exp{\left\{- i p(x-y)\right\}} \langle \psi(x) \bar{\psi}(z) \Gamma \psi(z) \bar{\psi}(y) \rangle\,.
\end{equation}
For a flavor singlet quark bilinear operator, the contraction of the quark fields of the vertex function gives rise to a connected
\begin{equation*}
 V^\Gamma (p)^{\textrm{conn.}} =   \sum_{x,y,z} \exp{\left\{- i p(x-y)\right\}} \langle \gamma_5 S(z,x)^\dag \gamma_5 \Gamma S(z,y) \rangle\,
\end{equation*}
and a disconnected contribution
\begin{equation*}
 V^\Gamma (p)^{\textrm{discon.}} =  - N_f \sum_{x,y,z} \exp{\left\{- i p(x-y)\right\}} \langle S(x,y) \textrm{Tr}_0(\Gamma S(z,z))\rangle\,,
\end{equation*}
$S(x,y)$ being the quark propagator. The expression $\textrm{Tr}_0$ denotes the trace of the operator minus its vacuum expectation value,
\begin{equation*}
\textrm{Tr}_0(\Gamma S(z,z)) = \textrm{Tr}(\Gamma S(z,z)) - \langle  \textrm{Tr}(\Gamma S(z,z)) \rangle\,;
\end{equation*}
the last term does not vanish for instance in the case of the operator $\Gamma = \mathds{1}$ and in other case the subtraction may reduce the statistical error.

The RI'-MOM scheme is a regularization invariant scheme, meaning that the condition (\ref{renocond}) can be consistently imposed both on the lattice and in the continuum quantum field theory employing the usual dimensional regularization. It can therefore provide a common scheme to eventually link lattice numerical results to the schemes commonly used in perturbation theory.

\section{Numerical calculation}
\begin{table}[t]
  \centering
\begin{tabular}{c|c|c|c}
                        & $\beta=5.20$ &  $\beta=5.29$ &  $\beta=5.40$ \\ \hline
$a [\textrm{fm}] $      & 0.081        &  0.071        &  0.060        \\ \hline
$r_0/a$                 & 5.454(20)    &  7.004(54)    &  8.285(74) \\\hline
$M_\pi [\textrm{MeV}] $ & 600-300      &  420-150      &  490-260
\end{tabular}
\caption{Summary of our $N_f=2$ ensembles used for the present calculation.}\label{ensable}
\end{table}
We perform the non-perturbative computation of the flavor singlet renormalization constants on the $N_f=2$ QCDSF-RQCD ensembles \cite{MG1,SB1,SB2}; pion masses and lattice spacings are summarized in Table~\ref{ensable}. We analyze 100 configurations for each ensemble separated by 10 MDUs. The vertex function is not a gauge invariant observable, therefore we measure $V$ on Landau gauge fixed configurations. The computation of the disconnected contribution to the vertex function is the most difficult task of the present study. We compute $\textrm{Tr}_0(\Gamma S(z,z))$ with 20 stochastic estimators and we improve the signal by means of the hopping parameter expansion. The momenta are chosen to have a symmetric direction in the dual lattice space.%; because our lattices have $2N_s = N_t$, we set
%\begin{equation}
%a p_n = 2 \pi \left\{\frac{n}{N_s},\frac{n}{N_s},\frac{n}{N_s},\frac{1}{N_t}\left(2n+\frac{1}{2}\right)\right\}\,.
%\end{equation}

\begin{figure}
\centering
 \subfigure[Tensor operator]{\includegraphics[width=.47\textwidth]{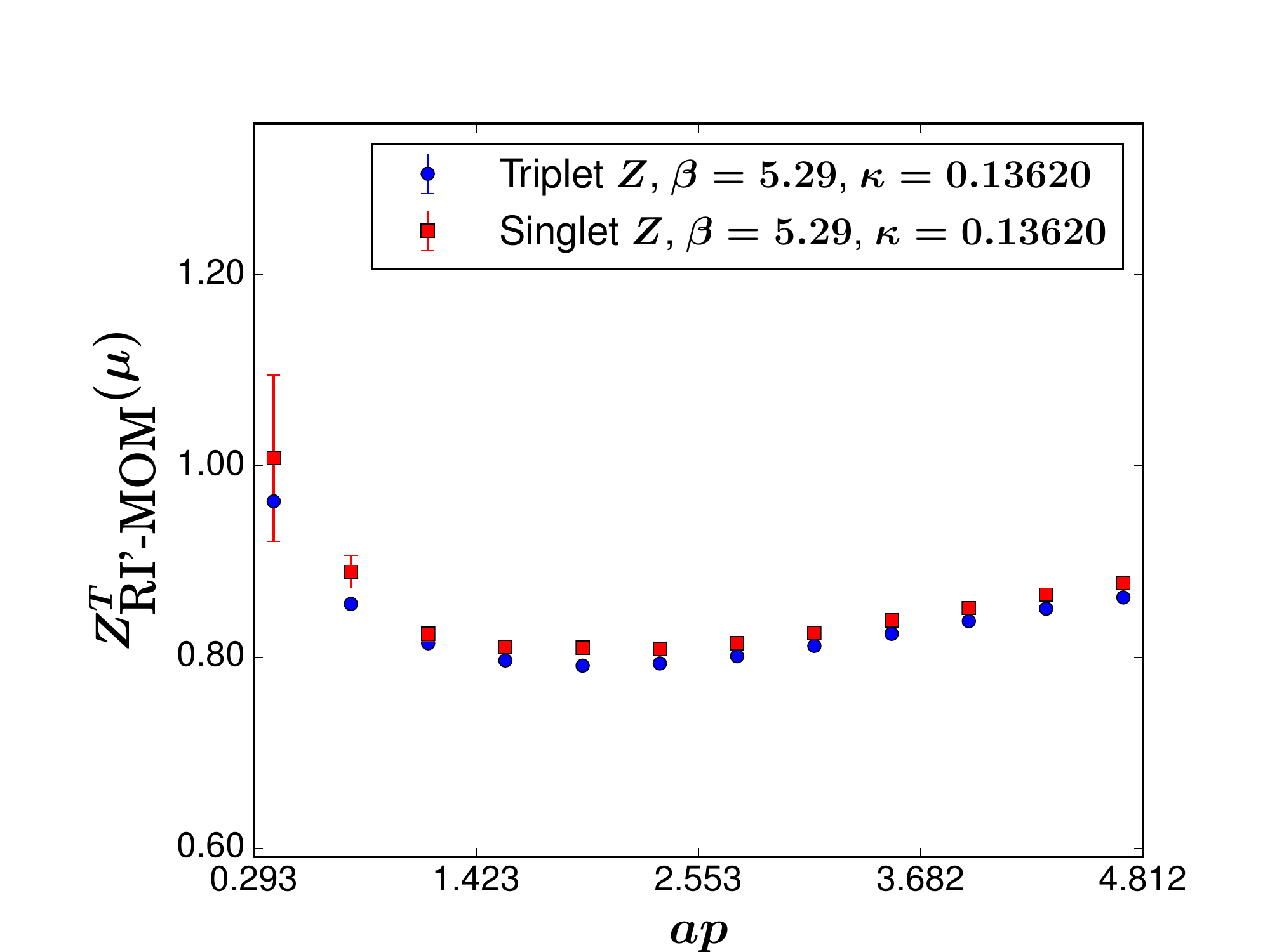}\label{tensor}}
 \subfigure[Axial vector operator]{\includegraphics[width=.47\textwidth]{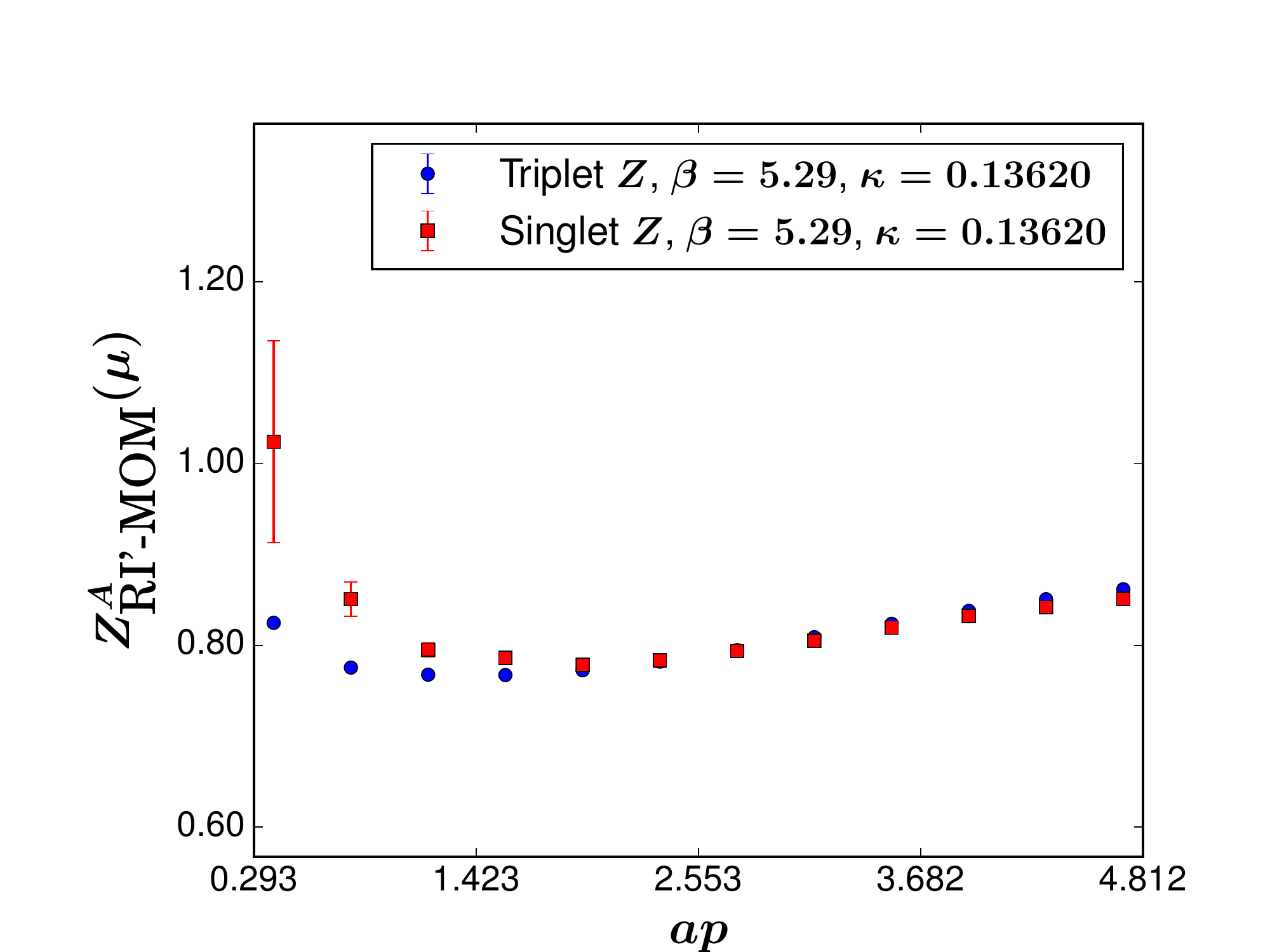}\label{axial}}
 \subfigure[Pseudoscalar operator]{\includegraphics[width=.47\textwidth]{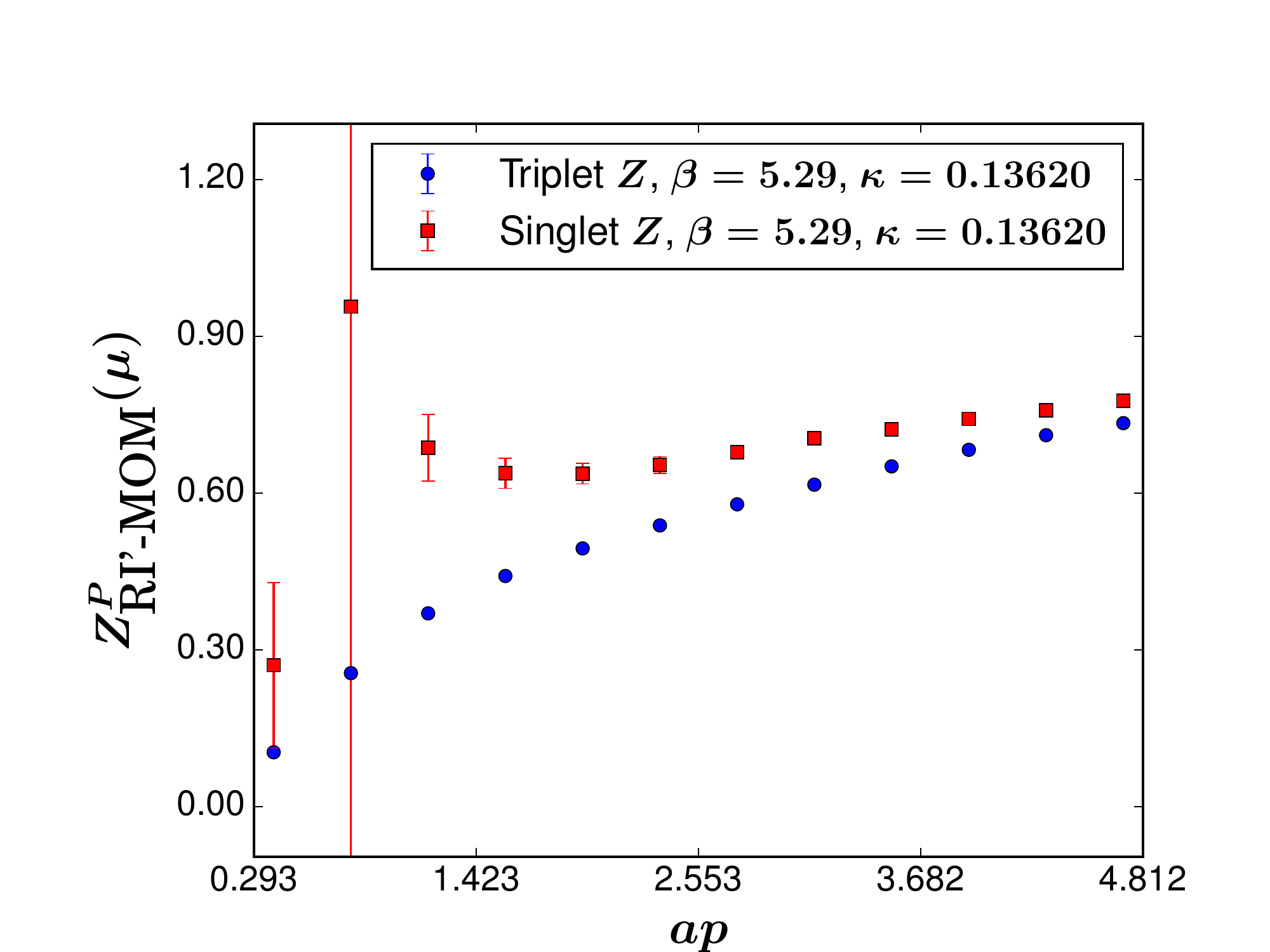}\label{pseudoscalar}}
 \subfigure[Scalar operator]{\includegraphics[width=.47\textwidth]{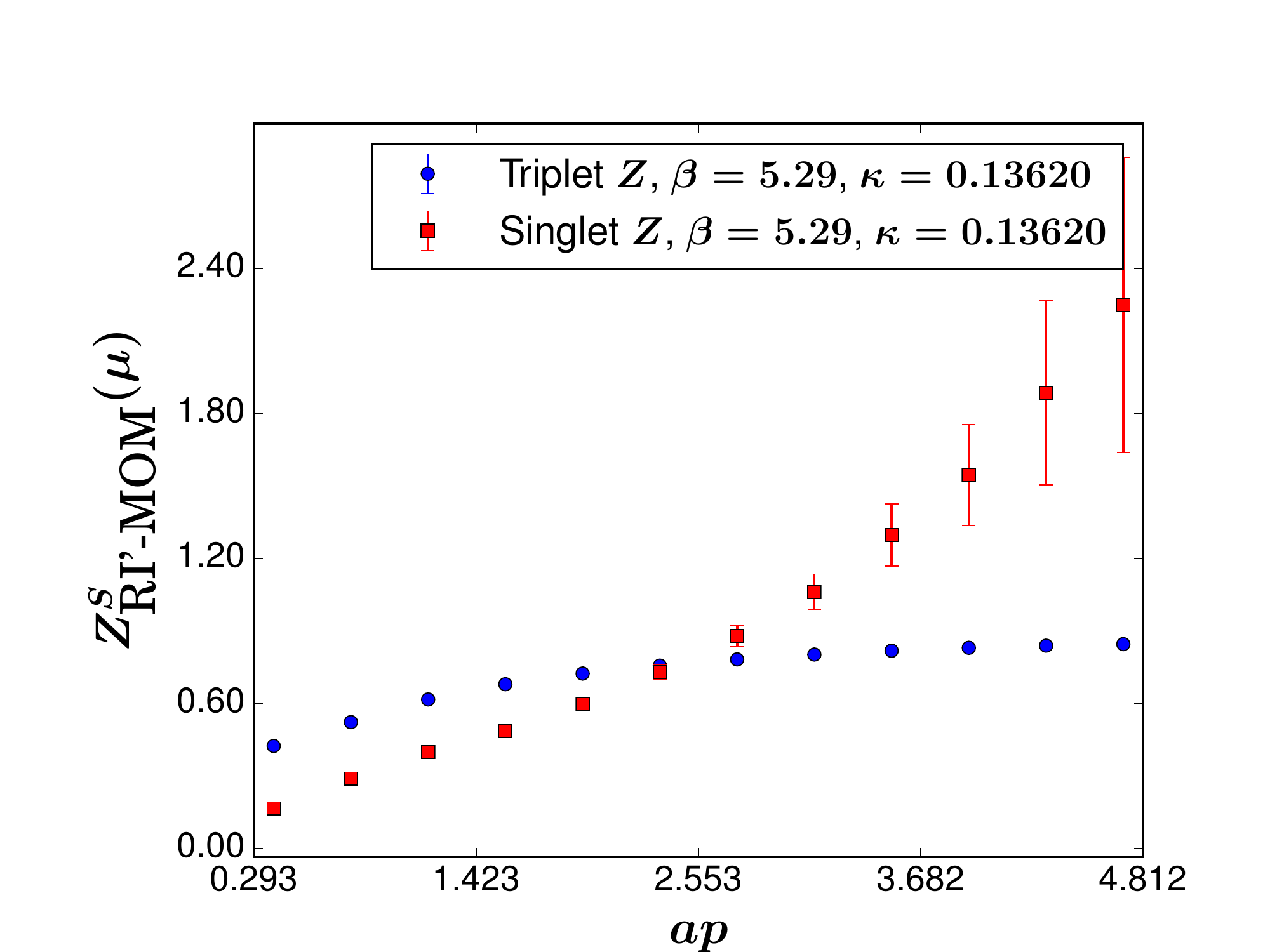}\label{scalar}}
 \caption{Renormalization constants in the RI'-MOM scheme for an ensemble with $\beta=5.29$ ($a \simeq 0.071$ fm) and $m_\pi = 420$ MeV for the a) tensor, b) axial vector c) pseudoscalar and d) scalar operator. The singlet renormalization constant differs only slightly from its triplet value for the tensor and axial vector operators.}\label{rimompct}
\end{figure}
We observe a small contribution coming from the disconnected contributions to the renormalization of the tensor $\bar{\psi} \sigma_{\mu\nu} \psi$ and axial vector $\bar{\psi}\gamma_5 \gamma_\mu \psi$ operators, see Fig.~\ref{tensor} and \ref{axial}. Larger corrections are observed for the pseudoscalar and scalar operator,  $\bar{\psi} \gamma_5 \psi$ and $\bar{\psi} \psi$. In the latter case there is evidence of strong lattice spacing effects at large momenta, see Fig.~\ref{pseudoscalar} and \ref{scalar}. Note that the singlet pseudoscalar operator is not expected to be affected by the pion pole, due to the $U_A(1)$ axial anomaly and the absence of a Goldstone boson in the singlet $0^{-+}$ channel. Therefore, the chiral extrapolation of $Z^P_{s.}$ is not singular.

\begin{figure}
\centering
 \subfigure[Tensor operator]{\includegraphics[width=.47\textwidth]{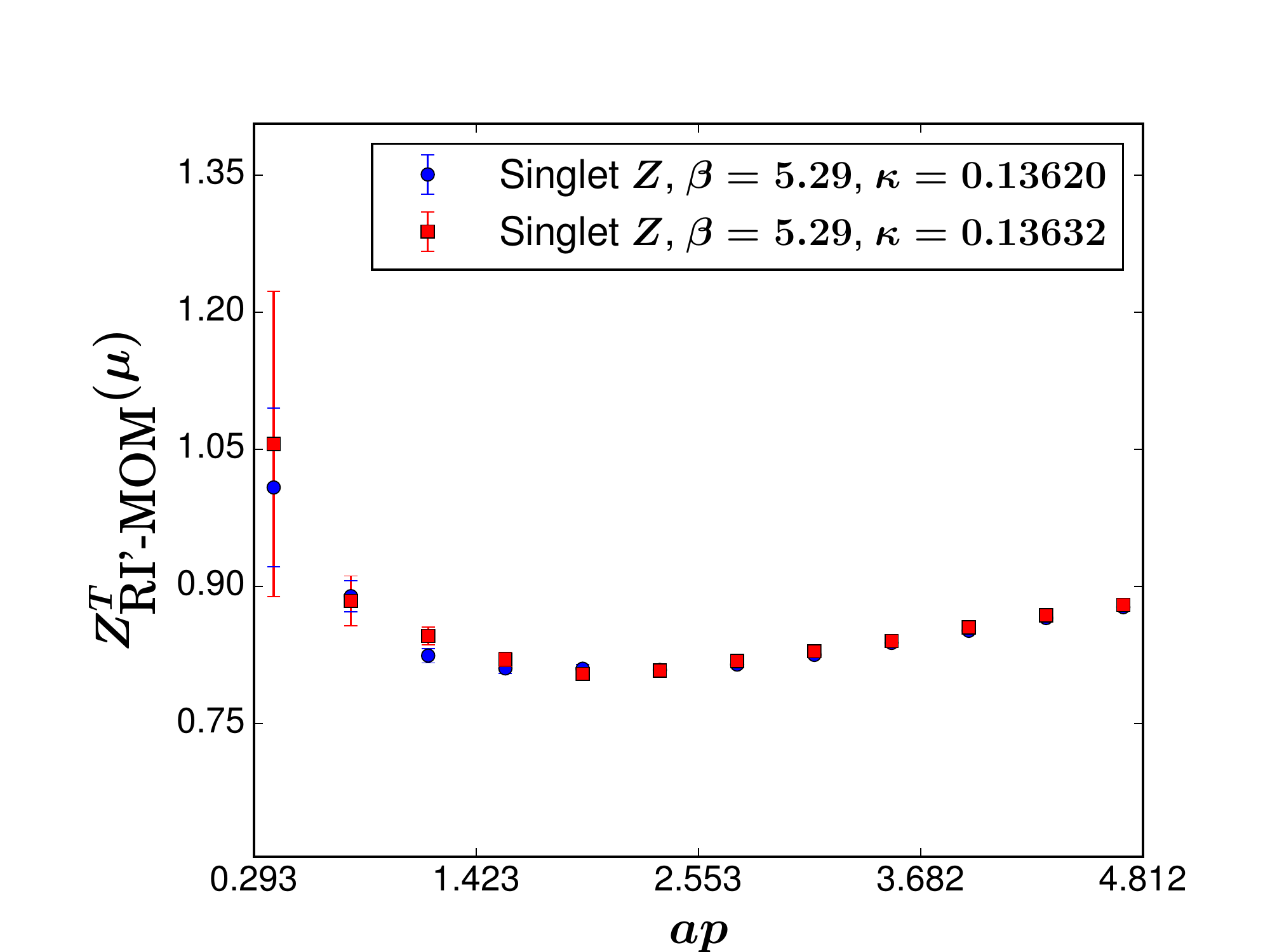}}
 \subfigure[Axial vector operator]{\includegraphics[width=.47\textwidth]{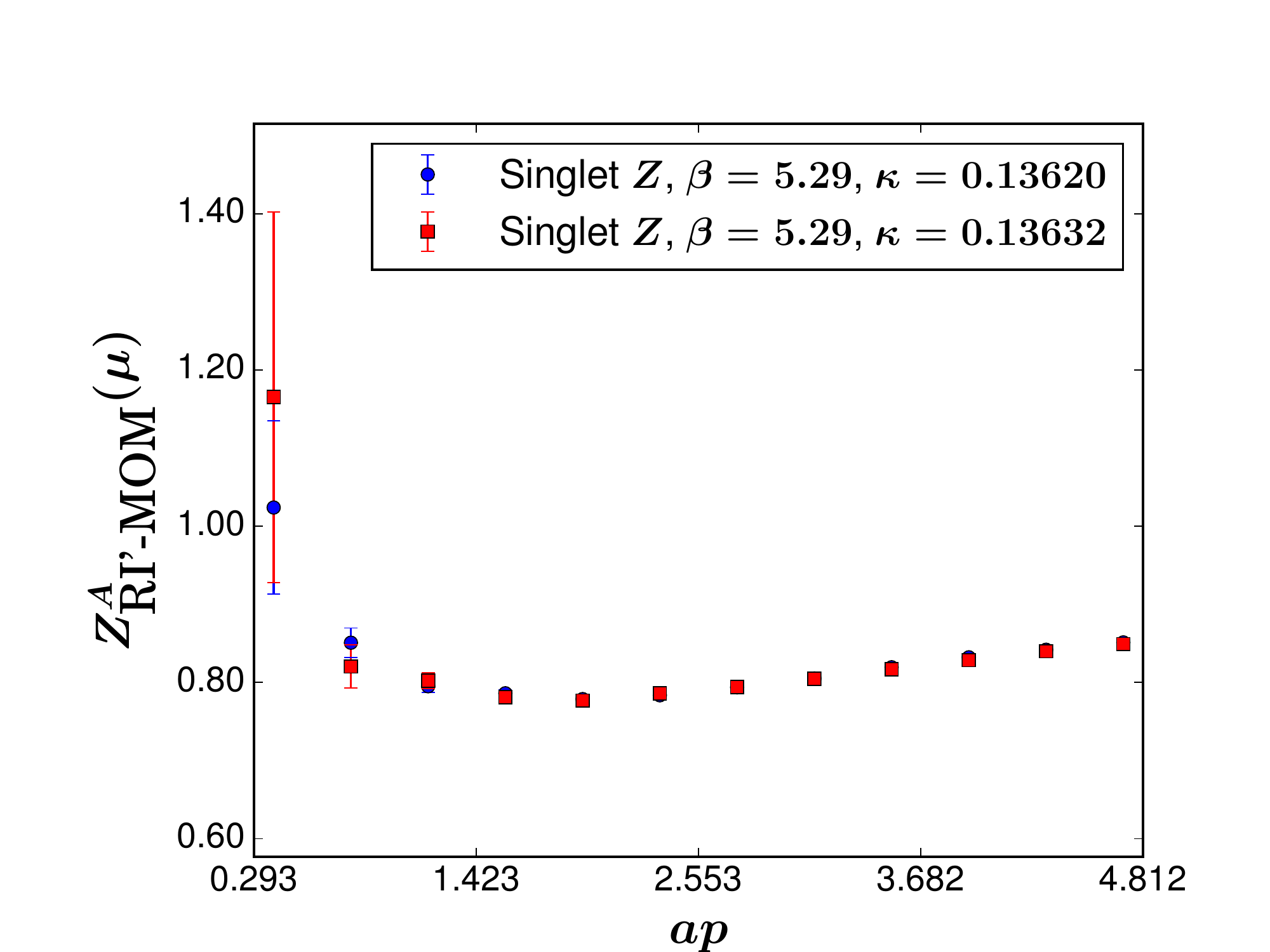}}
 \caption{Comparison of the axial vector and tensor singlet renormalization constant at pion mass $m_\pi~=~420$~MeV (blue points, $\kappa = 0.13620$) and $m_\pi = 295$ MeV (red points, $\kappa = 0.13632$).}\label{pionmasscomp}
\end{figure}
A comparison at different pion masses reveals a mild dependence of the flavor singlet renormalization constants on the quark mass, see Fig~\ref{pionmasscomp}.

\section{Matching to perturbation theory}

The non-perturbative calculation of the renormalization factors is presented in terms of the renormalization group invariants $Z^{\Gamma}_{\textrm{RGI}}$, to easily allow for a conversion of our results to a different renormalization scheme or to a different renormalization scale.

The $Z^{\Gamma}_{\textrm{RGI}}$ are computed in two steps:
\begin{enumerate}
 \item First, we perform the conversion to the $\overline{\textrm{MS}}$ scheme. To this end, the computation in perturbation theory of the conversion factor $Z_{\textrm{RI'-MOM}}^{\overline{\textrm{MS}}}(\mu)$ is required. The conversion factors for singlet quark bilinear operators with an even number of gamma matrices do not differ from the non-singlet case, while they are unknown beyond one-loop in the case of the vector and axial vector currents.
 \item In the second step, we absorb the dependence of $Z_{\overline{\textrm{MS}}}(\mu)$ on scale $\mu$ by integrating the renormalization group equations 
 \begin{equation}\label{aneq}
 \gamma^{\overline{\textrm{MS}}} = - \mu \frac{d}{d\mu}  \log{Z^{\overline{\textrm{MS}}}(\mu)}\,,
 \end{equation}
 that provide the scale dependence of the operator $\Gamma$ in the $\overline{\textrm{MS}}$ scheme. The anomalous dimension $\gamma^{\overline{\textrm{MS}}}$ for $O_\Gamma$ is known from continuum perturbation theory (see, e.g. Ref.~\cite{MG1}), for the axial vector singlet it is given in Ref.~\cite{LA1}. After integrating Eq.~(\ref{aneq}), we introduce
 \begin{equation}
 \Delta Z_{\overline{\textrm{MS}}}(\mu) = \left(2 \beta_0 \frac{g^{\overline{\textrm{MS}}}(\mu)^2}{16 \pi^2}\right)^{-\frac{\gamma_0}{2 \beta_0}} \exp{\left\{\int_0^{g^{\overline{\textrm{MS}}}(\mu)} \left( \frac{\gamma^{\overline{\textrm{MS}}}(g') }{\beta^{\overline{\textrm{MS}}}(g')} + \frac{\gamma_0}{\beta_0 g'}\right) \operatorname{d}g'\right\}}\,.
 \end{equation}
\end{enumerate}
Finally, $Z^{\Gamma}_{\textrm{RGI}}$ reads
\begin{equation}
 Z^{\Gamma}_{\textrm{RGI}}(a) = \Delta Z_{\overline{\textrm{MS}}}(\mu) Z_{\textrm{RI'-MOM}}^{\overline{\textrm{MS}}}(\mu, a) Z_{\textrm{RI'-MOM}}(\mu, a)\,.
\end{equation}
It depends on the lattice spacing but is independent of the scheme and scale (assuming that the perturbative calculation of the anomalous dimension and of the conversion factor is accurate enough in the relevant region of $\mu$).

\begin{figure}
\centering
\includegraphics[width=.67\textwidth]{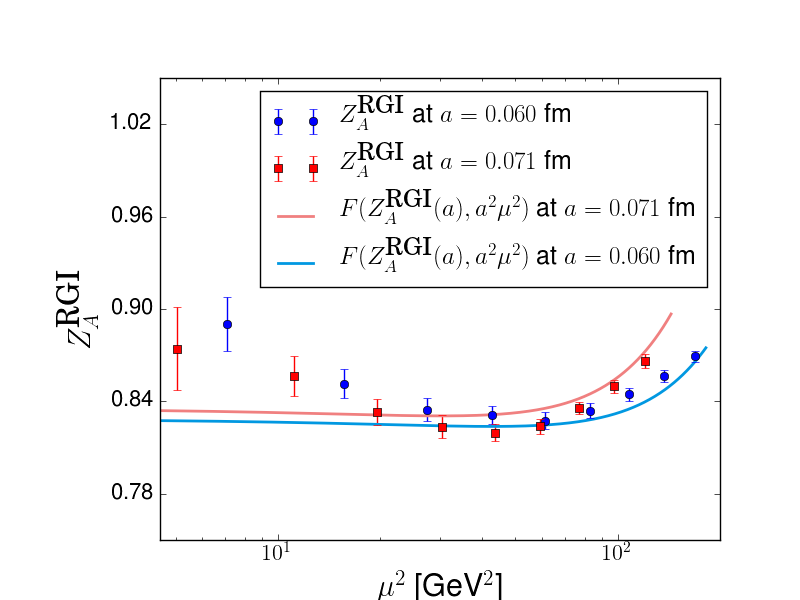}
\caption{Singlet renormalization group invariant $Z^A_{\textrm{RGI}}$ as a function of the scale $\mu^2=p^2$. Lattice artefacts are clearly visible at large $\mu \gtrsim 6$ GeV, while the lack of higher orders of perturbation theory is the source of the deviations at $\mu \lesssim 3$ GeV. The two lines are the result of a single combined fit to all points with $\mu^2 > 12$~GeV${}^2$.}\label{zargi}
\end{figure}
We use $r_0 \Lambda_{\overline{\textrm{MS}}} =  0.789(52)$ for the matching with perturbation theory and the computation of the conversion factors \cite{AL1}. As an example, the RGI singlet axial vector renormalization constant is presented in Fig.~\ref{zargi} for two ensembles at the same pion mass $m_\pi= 420$ MeV but at two different lattice spacings, $a = 0.071$ and $a = 0.060$. The singlet $Z^A_{\textrm{RGI}}$ shows a large deviation from a constant behavior especially at small and large momentum scales, due to the effects of higher order of perturbation theory in the expansion (\ref{aneq}) and of non-zero $a$ respectively. The so-called ``window problem'' refers to the difficulties in lattice Monte Carlo simulations to find a region of the scale $\mu$ where the systematic uncertainties related to both lattice artefacts and perturbation theory are under control. Finer lattice spacings enlarge the window where the RGI renormalization constants can be reliably extracted and reduce lattice artefacts at large momenta. 

\begin{table}[t]
  \centering
\begin{tabular}{c|c|c}
 Operator                 & Flavor singlet                       &  Flavor triplet    \\ \hline
$Z_{\textrm{RGI}}^A$      & $0.835 (~7) (\substack{10 \\ -20})$  &  0.77682(54)(6)(-60) \\ \hline
$Z_{\textrm{RGI}}^P$      & $0.421 (10) (\substack{8  \\ -10})$  &  0.36711(11)(4)(-1) \\ \hline
$Z_{\textrm{RGI}}^T$      & $0.896 (12) (\substack{10 \\ -4})$   &  0.9368(14)(-39)(0) \\ \hline
$Z_{\textrm{RGI}}^S$      & $0.282 (10) (\substack{20 \\ -13})$  &  0.45155(80)(568)(15)
\end{tabular}
\caption{Summary of our results of the RGI renormalization constants for flavor singlet and triplet operators at the pion mass $m_\pi~=~420$~MeV and at $a=0.060$ fm, the first quoted error is statistical and the second systematic. The triplet renormalization constants have been listed in Table V of Ref.~\cite{MG1} and are quoted here for comparison.}\label{zargi420}
\end{table}
A possible solution to the problem of the lattice artefacts is a combined fit of the renormalization constants computed from \emph{all} different $a$ following the ansatz
\begin{equation*}
 F(Z_{\textrm{RGI}},a^2 \mu^2) = Z_{\textrm{RGI}}(a) + c_1 a^2 \mu^2 + c_2 (a^2 \mu^2)^2 + \dots \,.
\end{equation*}
The various choices for the range of $\mu^2$ to consider in the fit and the order of the polynomial are included in the final result as systematic uncertainties. 
The RGI renormalization constants computed following this method from the data at $m_\pi = 420$ MeV at $a=0.060$ fm are presented in Table~\ref{zargi420}. Despite the larger statistical errors compared to the triplet case, the even larger systematic uncertainties dominates the final error of the flavor singlet renormalization constants. Our results are in agreement with the behavior for the singlet renormalization constants $Z^A$ and $Z^S$ of Ref.~\cite{CH1}, although a direct comparison is not possible due to the different number of quark flavors of our simulations.

\section{Conclusions}

We presented the first calculation of the renormalization constants of flavor singlet quark bilinear operators for the $N_f = 2$ QCD theory.
The computational cost of the non-perturbative renormalization of flavor singlet quark bilinear operator is approximatively $O(10)$ times larger compared to the flavor non-singlet case. Our results show that it is possible to perform the calculation of the disconnected contribution to the vertex function using standard stochastic noise techniques. 

The analysis of a third ensemble at a lighter pion mass and the extrapolation of the renormalization constants to the chiral limit is our next step to conclude the analysis of the $N_f = 2$ flavor theory.
In the future, we plan to extend our analysis to the $N_f = 3$ CLS ensembles and we plan to include the operators relevant for the renormalization of the stress-energy  tensor.

\section{Acknowledgements}

This work was supported by the Deutsche Forschungs-gemeinschaft Grant  No.  SFB/TRR  55. The authors gratefully acknowledge the computing time granted by the Leibniz-Rechenzentrum (LRZ) in M\"unchen provided on SuperMUC. Part of the analysis  was  performed  on  the SFB/TRR~55  QPACE  2 Xeon-Phi installation in Regensburg.

\end{document}